# Opportunities and challenges for spintronics in the microelectronic industry

B.Dieny[1], I.L.Prejbeanu[1], K.Garello[2], P.Gambardella[3], P.Freitas[4,5], R.Lehndorff[6], W.Raberg[7], U.Ebels[1], S.O.Demokritov[8], J.Akerman[9,10], A.Deac[11], P.Pirro[12], C.Adelmann[2], A.Anane[13], A.V.Chumak[12,14], A.Hiroata[15], S.Mangin[16], M.Cengiz Onbaşlı[17], M.d'Aquino[18], G.Prenat[1], G.Finocchio[19], L.Lopez Diaz[20], R.Chantrell[21], O.Chubykalo-Fesenko[22], P.Bortolotti[13]

1. Univ. Grenoble Alpes, CEA, CNRS, Grenoble INP, IRIG, SPINTEC, Grenoble, France
2. Imec, Leuven, Belgium
3. Laboratory for Magnetism and Interface Physics, Department of Materials, ETH Zurich, Zurich, Switzerland.
4. International Iberian Nanotechnology Laboratory (INL), Braga, Portugal
5. Instituto de Engenharia de Sistemas e Computadores-Microsistemas e Nanotecnologias (INESC MN), Lisboa, Portugal
6. Sensitec GmbH, Mainz, Germany
7. Infineon Technologies AG, Neubiberg, Germany
8. Institute for Applied Physics, University of Muenster, Muenster, Germany
9. Department of Physics, University of Gothenburg, Sweden
10. Department of Applied Physics, School of Engineering Sciences, KTH Royal Institute of Technology, Sweden
11. Helmholtz-Zentrum Dresden—Rossendorf, Institute of Ion Beam Physics and Materials Research, Dresden, Germany
12. Fachbereich Physik and Landesforschungszentrum OPTIMAS, Technische Universität Kaiserslautern, Kaiserslautern, Germany
13. Unité Mixte de Physique CNRS-Thales, Univ. Paris-Sud, Univ. Paris-Saclay, Palaiseau, France
14. Faculty of Physics, University of Vienna, Vienna, Austria
15. Department of Electronic Engineering, University of York, Heslington, United Kingdom
16. Institut Jean Lamour, Université de Lorraine, Nancy, France
17. Koc University, Istanbul, Turkey
18. Department of Engineering, University of Naples "Parthenope", Naples, Italy
19. Department of Mathematical and Computer Sciences, Physical Sciences and Earth Sciences, University of Messina, Messina, Italy
20. Universidad de Salamanca, Department of Applied Physics, Salamanca, Spain
21. Department of Physics, The University of York, Heslington, United Kingdom
22. Instituto de Ciencia de Materiales de Madrid, CSIC, Spain

Contact emails: bernard.dieny@cea.fr; paolo.bortolotti@thalesgroup.com

**Spin-based electronics has evolved into a major field of research that broadly encompasses different classes of materials, magnetic systems, and devices. This review describes recent advances in spintronics that have the potential to impact key areas of information technology and microelectronics. We identify four main axes of research: nonvolatile memories, magnetic sensors, microwave devices, and beyond-CMOS logic. We discuss state-of-the-art developments in these areas as well as opportunities and challenges that will have to be met, both at the device and system level, in order to integrate novel spintronic functionalities and materials in mainstream microelectronic platforms.**

Conventional information processing and communication devices work by controlling the flow of electric charges in integrated circuits. Such circuits are based on nonmagnetic semiconductors, in which the electrons' spin does not play a role. Spintronic devices, on the other hand, use the spin degree of freedom to generate and control charge currents as well as to interconvert electrical and magnetic signals. By combining processing, storage, sensing, and logic within a single integrated platform, spintronics can complement and, in some cases, outperform semiconductor-based electronics. Indeed, the seamless integration of these functions can bring enormous benefits in terms of scaling, power consumption, and data processing speed of integrated circuits.[1]

In the thirty years that followed the Nobel Prize winning discovery of giant magnetoresistance[2,3] (GMR), which kick-started the field, an impressive amount of scientific and technological knowledge has been accumulated. Fundamental research delivered insight into the behavior of nanostructured magnetic systems under the action of electric currents[4,5,6], the interplay of charge and spin transport phenomena[7,8], and the dynamics of spins and spin excitations in magnetic[9,10,11] as well as nonmagnetic[12,13] systems. In parallel to such progress, GMR-based spin valves and magnetic tunnel junctions (MTJs) rapidly found large-scale commercial applications as magnetic field sensors in tape and disk read heads, and as position or proximity sensors in cars, automated industrial tools, and biomedical devices. The discovery of spin-transfer torques[14,15] (STT), together with the optimization of the tunneling magnetoresistance (TMR) and magnetic anisotropy in MgO-based MTJs[16,17,18], further enabled the realization of scalable nonvolatile magnetic random access memories[19,20,21] (MRAMs). Owing to their low energy consumption, fast switching, and superior endurance, STT-MRAMs are presently commercialized as a replacement for SRAMs in embedded cache memories, with potential applications also as a persistent DRAM technology.

The sensing and storage capabilities of spin valves and MTJs exemplify the prototypical functions of spintronic devices. Both types of devices are based on thin film structures consisting of alternating ferromagnetic and nonmagnetic layers, in which the relative alignment of the magnetization in different elements of the stack influences the scattering or ballistic transmission of spin-polarized electrons, thus determining the electrical resistance of the device. The magnitude of the resistance is used to sense the orientation of one magnetic layer with respect to another, that is, to indicate whether a "1" or "0" state is stored in the parallel or antiparallel alignment of the two ferromagnetic layers. The same current that is employed to read the magnetic state of the device can be used to switch it, if the current density is increased beyond a critical threshold, by transferring spin angular momentum from the so-called reference layer to the free layer via the STT. Thus, all electrical operation is achieved, with a tunable trade-off between high speed and low power[22].

Technologies based on GMR and MTJ devices are now firmly established and compatible with CMOS fab processes. Yet, in order to meet the increasing demand for high-speed, high-density, and low power electronic components, the design of materials, processes, and spintronic circuits needs to be continuously innovated. Further, recent breakthroughs in basic research brought forward novel phenomena that allow for the generation and interconversion of charge, spin, heat, and optical signals. Many of these phenomena are based on non-equilibrium spin-orbit interaction effects, such as the spin Hall and Rashba-Edelstein effects[6,8,23] or their thermal[24] and optical[25,26] analogues. Spin-orbit torques (SOT), for example, can excite any type of magnetic materials, ranging from metals to semiconductors and insulators, in both ferromagnetic and antiferromagnetic configurations[6]. This versatility allows for the switching of single layer ferromagnets, ferrimagnets, and antiferromagnets, as well as for the excitation of spin waves and auto-oscillations in both planar and vertical device geometries[10,11]. Charge-spin conversion effects open novel pathways for information processing using Boolean logic, as well as promising avenues for implementing unconventional neuromorphic[27,28,29] and probabilistic[30] computing schemes. Finally, spintronic devices cover a broad bandwidth ranging from DC to THz[31,32], leading to exciting opportunities for the on-chip generation and detection of high frequency signals.

Following these developments, this review focuses on spintronic phenomena and applications that have demonstrated proof-of-concept devices, corresponding to a technology readiness level (TRL) ≥ 3, and have a strong potential for medium-term implementation in integrated or embedded electronic systems. The aim of the review is two-fold: on the one hand, we wish to highlight present and future assets of spintronics for the broader electronic engineering community; on the other hand, we wish to encourage the spintronics community to "think" applications, with the goal of translating more basic research into industrial technologies and economical gains. The review is organized around four axes of research and development, as identified by SpintronicFactory[33], a pan-European network that federates major academic and industrial actors in the field of spintronics: 1) Memories, 2) Magnetic sensors, 3) Radio frequency and microwaves devices, 4) Logic and non-Boolean devices. These axes are supported by two transverse activities: A) Advanced materials, nanofabrication and tests, B) Modelling and design.

**Magnetic memories**

The demand for on-chip memories is constantly increasing due the exponential growth of data storage requirements and a rising gap between processor and off-chip memory speeds. Energy consumption at the chip-level has also increased exponentially, reaching a power-wall for the safe operation of devices. Volatile memory components greatly contribute to the power budget of integrated circuits. One of the best solutions to limit power consumption and to fill the memory gap is the modification of the memory hierarchy by integration of non-volatility at different levels, which would minimize static power and also enable normally-off / instant-on computing. Among the proposed nonvolatile memories, magnetoresistive MRAMs are particularly attractive, owing to their low-voltage and high-speed operation capability in addition to their high-endurance.

*Current status.* MRAM technologies evolved in the last years, benefiting from discoveries in spintronics, namely the tunnel magnetoresistance (TMR) of magnetic tunnel junctions (MTJ)[16-17] which enables large read-out signals, and the spin transfer torque (STT)[14,15] and spin-orbit torque (SOT) phenomena [34], which enable electrical switching. STT-MRAM (Fig.1b) attracted much attention owing to their low-power consumption and high performance. Driven by scalability, and taking advantage of the perpendicular magnetic anisotropy at the CoFeB/MgO interface[18], industrial efforts focus on perpendicularly magnetized STT-MRAM as it offers large retention and small cell footprint. Leading semiconductor industries and tool suppliers have aggressively launched their development program for STT-MRAM for either eFlash or SRAM replacement[35-38]; recently Intel announced NOR-Flash replacement, while Samsung and Everspin/GF announced the release of a 1GB embedded MRAM on the 28 nm node.

*Future challenges: density.* In order to promote MRAM as a viable solution for DRAM replacement, several solutions for the patterning of MTJ at very narrow pitch have been proposed, either by improved etching chemistries or by unconventional strategies, such as depositing the MTJ on pre-patterned pillars[39]. A new concept to increase the downsizing, called Perpendicular Shape Anisotropy PSA-STT-MRAM[40], has been introduced, which consists in significantly increasing the thickness of the storage layer so as to induce a perpendicular shape anisotropy that reinforces the interfacial anisotropy. Hence, large thermal stability factors can be achieved down to sub-10 nm diameters. Researchers are also very active in the development of storage shift register devices called "race track memories"[41]. The information is stored in the form of tracks of polarized spin textures (domain walls or skyrmions[42]) which can be moved by an electrical current along the tracks – Figure 1b. Such storage devices could reach extremely high storage densities without mechanical motions.

*Future challenges: power and speed.* The aim is to lower the STT current below 2 MA.cm$^{-2}$ and increase the TMR window for faster reading while keeping long retention times. Optimized stacks that comprise two tunnel barriers with antiparallel polarizing layers maximize anisotropy, retention and STT efficiency have been proposed and demonstrated[43].

A new class of magnetic memory called spin-orbit-torque MRAM (SOT-MRAM)[34,44]– Figure 1.c and 1.d – offers cache-compatible high speed and improved endurance, at the cost of larger writing currents and footprint. In order to reduce the writing current and therefore the size of the selection transistor, charge-spin conversion materials with low resistivity and large SOT efficiency must be developed. To be operated in stand-alone mode, these devices require applying a static magnetic field bias, for which several embedded solutions have been recently proposed, some demonstrated on large scale wafer[44].

Since most of the power dissipation is due to Ohmic conduction, electric field control of magnetism has been suggested as a new MRAM writing mechanism, enabling significant reduction of the energy consumption[45-47]. As no current is needed to operate the device except during the read operation, the power consumption is reduced to the charging/discharging energy through the MTJ, which results in a very low switching energy compared with that of STT switching. Although significant results have been achieved in voltage-controlled MRAM, there are still some issues for practical use. First, the voltage induced change of magnetic anisotropy (VCMA) with reasonable applied voltage (~1V) is so far too weak to be able to induce magnetization switching while insuring sufficient magnetic thermal stability in standby. Second, the writing time window is small and size dependent, resulting in reliability issues. Besides, like SOT, voltage control also requires an in-plane field. A promising perspective to mitigate these issues is offered by hybrid VCMA-STT combinations in 2-terminal devices or by VCMA-SOT combinations in 3-terminal MTJ devices, which enable write speed acceleration, lower current thresholds, as well as selective SOT switching of several MTJs sharing a single write line[47].

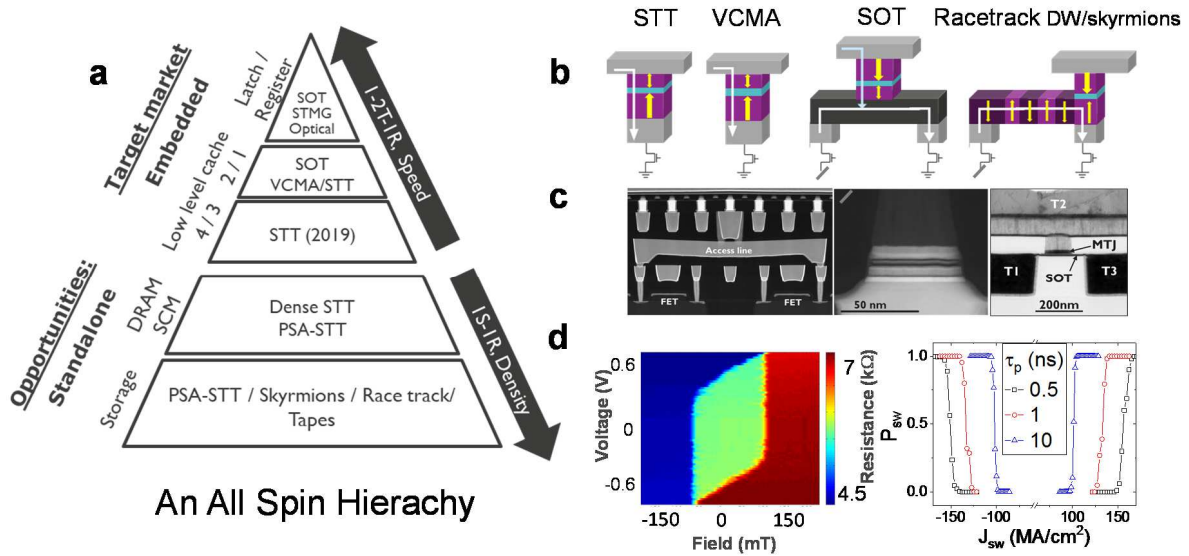

**Fig. 1| Nonvolatile magnetic memories.** (a) Proposed modification of the Memory hierarchy by spintronic solutions. (b) Various MRAM architectures (as described in the text). (c) TEM cross section of STT and SOT memory cells (courtesy of IMEC). (d) Left: classical STT-MRAM switching diagram (courtesy of SPINTEC), right: probability of switching as a function of the injected current for a 50nm SOT-MRAM memory cell (courtesy of IMEC).

**Magnetic sensors**

Spintronics based magnetic sensor technology has seen a rather strong development in the past 10 to 15 years with the advent of TMR sensors, gradually replacing GMR, AMR, and conventional Hall effect technologies in applications where higher output and signal to noise ratio (SNR), good thermal stability, compatibility with CMOS integration, reduced cost and minimum features (< 1 mm$^2$) are required[48-53]. Major markets today include the automotive sector (angular, speed, current, position/switch sensors), industry 4.0 with current and power sensors, linear and angular encoders, scanners, and consumer electronics / smart phones (3D magnetometers/digital compasses). Novel applications are emerging in the IoT and biomedical arenas with the development of low consumption, reliable devices on silicon or flexible substrates[48,50].

*Current status.* A typical MgO-based TMR sensor stack consists of Buffer/AF/SAF/MgO/FL/cap[50] where AF is the antiferromagnetic layer exchange biasing the pinned layer in the synthetic antiferromagnet (SAF). The free layer (FL) on MgO can be prepared with linear ranges from few mT to hundreds of mT. TMR values reach typically 200% and Resistance x Area (RA) products (from few hundred Ohm.µm$^2$ to several kOhm.µm$^2$) can be tuned for a variety of sensing applications by playing with the thickness of the MgO barrier. Temperature coefficients for TMR range from 500 ppm/°C to few thousand ppm/°C. MR sensors can be used in a very wide range of frequencies, from DC to hundreds of MHz. 1/f noise determines the minimum field detectivity at low frequency, leading to detectivities of few hundred pT at DC[52]. Hybrid architectures (magnetic flux guides) can lead to 10-20pT at 10Hz, and detectivities below 1pT in the white noise[53]. For sensor applications, null outputs are required for zero external excitation, requiring MTJ sensor integration in a Wheatstone bridge. Figure 2 shows an example of TMR-based probe system used to detect defects in metal surfaces.

In the automotive industry, solutions based on various magnetic sensor principles are used for example in ABS (anti blocking systems), drive by wire, engine management and ESP (electronic stabilization program). Hall-effect based devices are dominant today, but over the last 10 years, the transition to magnetoresistive sensors based on GMR and recently TMR has started and the number of products using spintronics phenomena steadily increases. This transition is triggered by increasingly stringent requirements for magnetic sensors accuracy and sensitivity required for advanced driver assistance systems and autonomous driving solutions. For these high volume applications, TMR sensors are monolithically integrated onto CMOS (as in MRAM). In a broader industrial environment (industry 4.0), spintronic sensors are being used in a variety of applications such as low volume highly specialized measurement tasks (scanners) to medium volume, standard applications like linear or angular encoders, and current sensors. Today, AMR and GMR sensors are widely established in industrial applications providing high resolution and robustness in adverse conditions in comparison to competing optical or Hall effect based encoders. TMR devices are stepping in holding advantages in resolution, temperature and lifetime drift. From a packaging point of view, the current standard is the integration of magnetoresistive sensors with signal conditioning and higher-level electronics in discrete packages on PCBs. Only in medium volume scenarios, the sensors are side-by-side integrated with application specific integrated circuits (ASICs).

Emerging applications in the biomedical area include spintronic biochip platforms where bioanalytes are labelled with magnetic nanoparticles, and where magnetoresistive sensors (mostly GMR based) detect their hybridization to immobilized probes [48-50]. These platforms have been used for proteins, or DNA based recognition. Another type of MR-based biochip is the integrated cytometer where labelled cells/bacteria are detected on the fly as they pass on microfluidic channels with incorporated MR sensors[48-50].

*Future challenges.* In order to improve the SNR, materials development will aim at reaching larger TMR values. Proper growth control is required to achieve good lattice matching between the different active layers using standard substrates with appropriate buffers and industry-compatible sputter deposition methods. Perpendicular magnetized materials may be used for larger dynamic ranges[54]. This could be also the playground for integrated 3D magnetometers. Linear sensors use single domain free layer state leading to hysteresis free response for fields applied in the hard axis direction. Novel vortex based sensors[55] offer an alternative solution with a topologically protected spin configuration that deserves further investigation. Control of magnetic contribution to 1/f noise and its link to easy axis dispersion and magnetization fluctuations still need further research[56,57], since this is at present a limiting factor for sensor detectivity. On a longer term novel magnetic sensing mechanisms will be exploited.

In the automotive market, significant changes will occur in the coming years to make cars (or other vehicles) cleaner, safer and smarter. In the near term, a very accurate control of the combustion engine and an improvement of hybrid vehicles efficiency are necessary. Spintronic sensors can help if their accuracy can be guaranteed over the vehicle´s lifetime under demanding environmental conditions and at a competitive cost at system level. On a longer term, the full electrification of vehicles and other transportation systems will increase the demand for magnetic sensors in all automotive applications. In comparison with combustion engines, an even higher accuracy of magnetic sensors will be required to control the drivetrain or powertrain of hybrid or fully electrified vehicles. This accuracy has to be ensured against the presence of magnetic disturbances originating from the electric motor and its periphery, which poses an additional challenge to the sensor system. Here, the performance of spintronic sensors is not yet sufficient to fully replace, e.g., shunt sensors for current measurements. Improved spintronic sensors will almost certainly play an important role in this process due to their high sensitivity and small size, which can be translated into high performance and low cost.

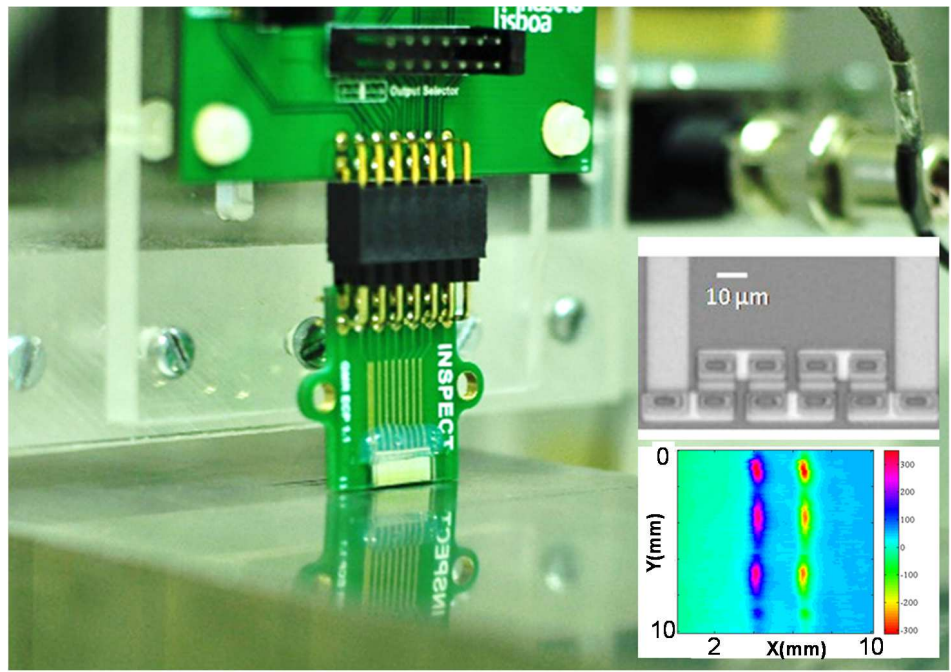

**Fig. 2| Magnetic sensors.** MTJ sensor array used for non-destructive testing of surface defects. Insets show one MTJ sensor channel and the surface scan signal. (Courtesy of INESC ID and INESC MN)

## Radiofrequency and Microwaves devices

Information and Communication Technologies (ICT) require besides sensing, storing and processing, the transmission of information using wireless communication systems. Common systems operate in the GHz range, where major markets are the Internet of Things (IoT) with bands in the 0.3 – 5.5 GHz range and public mobile networks (2G - 4G) using bands within 0.7 - 3GHz. There is a strong push (5G and beyond) to very high frequencies (hundreds of GHz towards THz) which will also be an enabler for future IoT[58]. Typical components for wireless communication are antennas, amplifiers, filters, delay lines, mixers, modulators and demodulators as well as frequency synthesizers (local oscillators) that have to be adapted to the corresponding frequency bands and communication protocols. For IoT more than 50 billion connected objects are expected by 2020[58]. This requires energy efficient, compact and low cost components and novel communication protocols in addition to defining novel approaches to process a large amount of data.

In this context, RF spintronics provides solutions for the most power-consuming parts of connected objects[59], which are the Transmitter (Tx)-Receiver (Rx) radio links, as well as for bio-inspired neuromorphic computing[60]. The combination of the oscillatory precession of magnetic moments with magnetoresistance phenomena enables two basic RF functions (see Fig. 3): DC-to-RF conversion for RF signal generation and RF-to-DC conversion for RF signal detection. These RF functions have been well established in the past for two basic device classes: (a) MTJ nanopillars exploiting the spin momentum transfer[14,15,61] and (b) two dimensional bilayer structures exploiting the inter-conversion of charge and spin currents[62-64] and the propagation of magnetic excitation (spin-waves or magnons). RF spintronic devices have the advantage of small footprint (~100nm) and can cover several decades of base frequencies between 0.1 and 60 GHz, reaching up to the THz range for antiferromagnetic materials. Moreover, these devices enable high modulation rates ($\geq$ 100 Mbs)[65] and provide multifunctionality at low power. RF devices thus complement and substantially extend the application range of spintronic devices.

*RF signal generation for wireless communication and neuromorphic computing.* RF signal generation exploits the DC-to-RF-signal conversion functionality within MTJ nanopillars (typical diameters of 50-500nm), see Fig. 3a(ii). Through the device configuration, different frequency ranges can be covered: the 0.1-1GHz range using vortex MTJs[61,62,66] or the 1-20 GHz range using quasi-uniform magnetized MTJs [59,65,67]. The nanoscale size, combined with the frequency tuning and output power levels of 1-10μW, make spintronic signal generators of interest for compact, broadband local oscillators used in transmitter and receiver modules. One target application is wireless communication within sensor networks for the 1-10 m range. First system level phase-locked loops[61] (PLL) have been developed in a hybrid CMOS/PCB technology demonstrating the capability of frequency synthesis at reduced phase noise (-90 dBm), see Fig. 3c. Future work will push PLL operation using advanced designs and concepts to improve phase noise and to demonstrate frequency shift keying protocols at data rates close to the PLL bandwidth (1-10 MHz). To push the data rate above 100 Mbs, spintronic devices offer compact solutions for amplitude[67], frequency[65] and phase shift keying (ASK, FSK, PSK) protocols. This is enabled through the non-linear coupling of amplitude and frequency where modulation can be achieved without the need of external mixers. First demonstrations show the potential of voice transmission over 10 m distances using conventional TV receivers, see Fig. 3c. Future efforts need to push the technology beyond TRL 3 by addressing two major issues. The first one is full integration within CMOS technology to demonstrate for instance RxTx modules, PSK communication and fast spectrum analysis. The second issue concerns reaching competitive performances by improving signal stability (phase noise), output power and achieving multifrequency operation. A major strategy is the implementation of networks of mutually synchronized devices[66,68]. The same networks of mutually synchronized MTJ devices also provide novel solutions for neuromorphic computing. Neurons in the brain form a network of coupled oscillators that can self-synchronize in frequency or phase. First experiments have demonstrated that a network of MTJ signal generators can be used to achieve recognition and classification operations by relying on the rich synchronization dynamics of the oscillators[60] (See Fig.4g). Much has yet to be understood on the physics of networks of mutual synchronized oscillating MTJ nano-devices.

*RF signal detection for power harvesting.* RF signal detection exploits the RF-to-DC signal conversion of MTJ nanopillars, see Fig. 3b(ii). Similar to signal generation, the device configuration determines the operational frequency range. DC output voltage levels in the µV to mV range have been demonstrated at input power levels as low as-60dBm[69]. This makes spintronic detectors competitive with Schottky diodes that have limited sensitivity at low input power levels. Furthermore, spintronic RF detectors are compact, frequency selective and tunable, and can be operated in two modes: (i) passive – without DC current bias and therefore at zero power consumption in the off-state and (ii) active – with DC current resulting in DC voltage levels 10-100 times higher than the passive mode. These properties open a large range of applications. The first example is wireless power harvesting, with a first demonstration of powering nanodevices using a broadband harvester[70]. The second example exploits the frequency filtering for ultra-low power and compact demodulators used within multifrequency protocols in wireless sensor networks. Future efforts need to push the technology beyond TRL 3 by integration within CMOS technology, by increasing the conversion efficiencies for energy harvesting as well as the SNR for signal demodulation and by defining suitable multifrequency protocols. As for signal generation, this passes by using networks of devices that are hence a central albeit non-trivial aspect of future RF spintronics research.

*Planar complex architectures.* The MTJ nanopillars use spin polarized charge currents to induce magnetization oscillations [Fig.3a(ii)]. The exploitation of charge-to-spin current conversion due to spin-orbit interactions in two dimensional ferromagnetic/heavy metal (FM/HM) bilayer systems [see Fig.3b(i)], opens many new alternatives not only for signal generation but also for localized generation of spin waves, for the development of filters and delay lines as well as for emission and detection in the THz range. For signal generation in the GHz range, recent experiments have demonstrated well defined steady state oscillations using single FM/HM nano-constrictions[63]. As compared to nanopillars, this configuration has the advantage of easier nanofabrication and synchronization of a large number of oscillators. Reduced linewidth and increased output power were demonstrated for an array of up to 8x8 synchronized nano-constrictions[64]. The future challenge lies in the efficient conversion of the magnetization dynamics into large electrical signals by combining the FM/HM devices with MTJs. This will open novel schemes for microwave signal generators and neuromorphic computing but also for developing electrically controlled efficient local spin wave emitters for magnonic applications that use spinwaves for low power information transport. These two dimensional device configurations offer many more opportunities when replacing the ferromagnetic FM material by insulating ferromagnets or by antiferromagnets. The strong exchange coupling in antiferromagnets pushes the corresponding frequencies into the THz range, giving prospect for the development of THz microwave emitters and detectors[71].

*RF signal processing by means of spin-waves.* Finally, using ferromagnetic insulating materials such as YIG[72] which has the lowest damping of known materials can provide novel approaches for various RF components based on spin-waves (also called magnons) needed in front end 5G communication systems. Magnons are elementary collective excitations of magnetic moments in magnetic materials and can be treated as quasiparticles. They have been known for about a century in solid-state physics but their emergence as potential innovative solutions in nanoelectronics has only been recognized quite recently.[10,11] The combination of magnons and nanomagnetism has led to a new, young field in spintronics research, named "magnonics". The growing interest about this research field is mainly stimulated by the new functionalities offered for RF applications by magnonic devices that are currently unavailable in conventional electronic devices or devices based on surface acoustic waves (SAW). Among them one could mention reconfigurability, scalability down to sub 100-nm sizes, wide operational frequency range from 1 GHz up to hundreds THz, nonlinear functionalities, etc. From the applications point of view, disruptive concepts of RF devices have been proposed including reconfigurable filters (e.g. based on so-called magnonic crystals), delay lines, phase shifters, frequency separators, Y-circulators and isolators, phase and amplitude multiplexers, wake-up receivers, spectrum analyzers, power limiters, signal-to-noise enhancers, wave front reversers, etc[10, 73, 74]. Most of these concepts are still investigated in laboratories at TRL 2 to 3. One of the main obstacles is the relatively low efficiency of the currently-available converters from RF electric signals to spin waves and back. As a consequence, insertion losses for micron-scale devices do not yet meet industry standards. Nevertheless, the constantly demanding miniaturisation of the RF components and the need for higher operational frequencies (e.g. due to 5G standard) and frequency tunability attracts increasing attention towards the spin-wave based technologies. Indeed this technology becomes more and more competitive in regards to alternative technologies as the operational frequency increases.

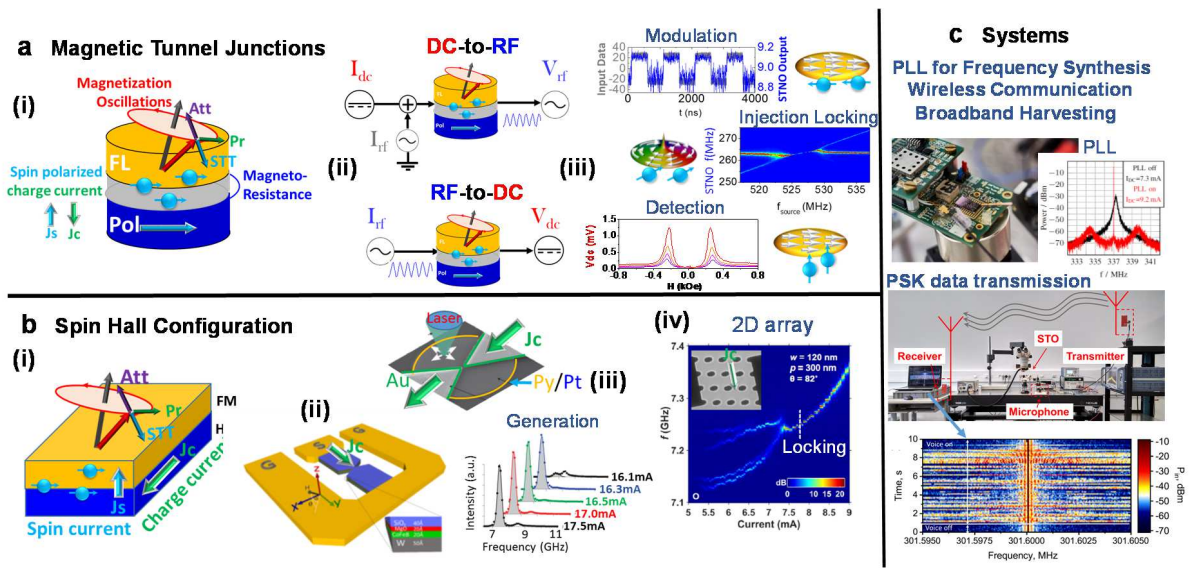

**Fig. 3| Radiofrequency and microwaves devices.** Basic RF Spintronics configurations are MTJ nanopillars a(i) and two dimensional FM/HM spin Hall bilayers b(i), in the form of nanoconstrictions b(ii,iii). Both configurations demonstrate DC-to-RF and RF-to-DC conversion functionality in the 0.1 – 20 GHz range and are illustrated for MTJs in a (ii). When the DC current reaches a threshold, signal generation sets in, as illustrated in b(iii)[62]. Adding an rf signal to the DC input, it is possible to modulate or lock by injection the generated signal, as illustrated in a(iii). Mutual coupling of several oscillators is an important concept to increase power and reduce linewidth, as demonstrated for arrays of nanoconstrictions in b(iv)[64]. Applying an rf current only, the resonance excitation leads to a rectified DC signal used for signal detection, as illustrated in a(iii). (c) First system level operations have been demonstrated, such as a hybrid CMOS/PCB PLL[61] and transmission of voice over 10 m distance using phase shift keying.

**Logic and non-Boolean devices**

The energy barrier that needs to be overcome to switch a nanomagnet is of the order of 0.1 aJ, two to three orders of magnitude lower than the energy per bit required for computation using conventional CMOS devices. There is therefore strong interest in using the magnetic degrees of freedom for ultralow power nonvolatile logic circuits[75]. Yet, despite intensive work during the last three decades, spintronic logic and computation is still at a rather conceptual level, with only few demonstrations reported beyond TRL 3. Initial research has focused on implementing the spin degrees of freedoms in analogues to conventional semiconductor devices[76]. Such devices, however, are based on dilute magnetic semiconductors with Curie temperatures below room temperature.

Besides semiconductor-based devices, numerous spintronic concepts for digital Boolean logic have been reported based on spin currents,[77,78] nanomagnets (Fig. 4a)[79-81], magneto-electric spin-orbit (MESO) logic (Fig. 4b)[82], exchange-driven magnetic logic (Fig. 4c)[81], domain walls,[83-85] skyrmions,[86] and spin waves (Figs. 4d, 4e)[87-90]. Spin currents can replace charge currents in logic switches and gates,[77,78] and information can be encoded in the polarization of the spin current. Potentially, large reductions of energy dissipation can be obtained if spin currents are decoupled from charge currents. Materials with weak spin-orbit interaction such as graphene provide sufficiently long spin lifetimes at room temperature to make these devices practical,[91] although an experimental demonstration of logic operation is still lacking.

Several other concepts are based on the coding of information in the magnetization state of a nanomagnet, providing a pathway to nonvolatile logic[79-81]. Computation can then occur via spin currents[78], dipolar[79,81] or exchange[81] coupling between nanomagnets. While robust room temperature operation has been demonstrated, the scalability as well as the energy-delay product of these approaches still need to be assessed on a circuit level. Recently, a nanomagnetic logic concept based on magnetoelectric control of a nanomagnet in combination with a readout scheme using the inverse spin-Hall effect was reported[82] (Fig. 4b). This concept promises ultralow power operation of circuits if the magnetoelectric coupling and the inverse spin-Hall effect are sufficiently large. However, the experimental demonstration of such a device is still lacking. Domain wall logic is a related concept that replaces individual nanomagnets by domains in extended ferromagnetic conduits[84] similar to the magnetic racetrack memory concept [41,92]. Here, the main challenge is the controlled and rapid domain wall propagation in nanoscale magnetic conduits at low power.

The field of magnonics offers a different approach to spintronic logic where information is coded in the amplitude or phase of spin waves.[10] Spin waves can have wavelengths down to the nm range and frequencies from GHz to THz, which renders them interesting for nanoscale fast logic technologies. Spin waves interference can be used for computation and provides a direct route towards majority gates.[88] Recently, a prototype of such a spin wave majority gate has been experimentally demonstrated (Fig. 4d).[90] While spin waves can have intrinsically low power, the current main limitation for such concepts is the lack of energy-efficient converters from spin waves to electric current and *vice versa*. All-magnon data processing requires a smaller number of converters but relies on non-linear magnonic effects (Fig. 4e)[87,88,93]. Experimental demonstrations of an integrated magnonic circuit[93], as well as an analysis at the circuit level, are still lacking although the critical building block - nanosized spin-wave directional coupler - was already reported recently.[88]

Another foundational concept, at the origin of a large number of discoveries is the three-terminal device imagined by Datta and Das[77] and named Spin Field-Effect Transistor (Spin-FET)[94]. Spin-FETs are composed of two ferromagnetic contacts playing the role of polarizer and analyzer (spin-polarized source and drain), separated by a gated semiconductor channel. Because of the non-zero spin-orbit interaction, the spin polarization at the drain is controlled by the voltage applied to the gate and electrons with spin aligned to the drain magnetization have higher probability to leave the channel than those with spin anti-aligned thus realizing on/off functionality[77,94]. This device was initially predicted to work faster and more efficiently than conventional FET with additional advantages like non-volatility of data and less heat generation. Even though a large number of material combination has been tried, this concept has not yet been validated for efficient room temperature operation due to the low energy efficiency of the spin-to-charge conversion at the semiconducting-ferromagnet interfaces. A significant breakthrough was however reported recently for the ballistic transport regime (at low temperature) in Ref. 95, where an all-electric and all-semiconductor Spin-FET utilizing two quantum point contacts as spin injectors and detectors was demonstrated.

Beyond Boolean logic, spintronics also offers many promising paths to alternative or unconventional computing schemes. Spintronics is attractive for ultralow power neuromorphic computation using spin waves (Fig. 4f),[96] spin-torque oscillators (Fig. 4g),[60] MTJ-based memristors (Fig. 4h),[97] or reservoir computing schemes (Fig. 4i).[98] Spintronics may also be an important part of quantum computation schemes, which is however beyond the scope of this review.

*Future challenges.* Most spintronic concepts describe fundamental building blocks of logic circuits that are equivalent to, *e.g.*, a transistor or a logic gate in conventional electronics. For example, spintronic majority gates have been realized, since they promise considerable circuit simplification and are not efficiently realized in CMOS.[83] Together with an inverter, majority gates are universal and can be used to synthesize any logic function.

For real computing applications, such fundamental building blocks need to fulfill a set of criteria, such as cascading, fan-out, logic level restoration, immunity to noise and losses, as well as input-output isolation. Only few concepts currently fulfill all these criteria[82] and can therefore be used to design more complex logic circuits. Moreover, there is currently no concept for a full spintronic computer, which incorporates logic, memory, and interconnects using exclusively magnetic signals without intermediate conversion to charge signals. Thus, real spintronic computing systems will need to be hybrids combining spintronic subsystems with CMOS electronics, in which spin signals need to be converted to charge signals and *vice versa*. The overall performance of the system in terms of energy consumption as well as computing throughput will depend crucially on the conversion mechanisms and the number of interconversions needed to perform the computation. Various physical phenomena including magnetoresistance, spin-transfer- or spin-orbit-torques, spin-Hall-effect, as well as inductive conversion can be used; however, their energy conversion efficiencies in nanoscale devices are rather low and the energy consumed in the CMOS "periphery" typically outweighs the *intrinsic* energy savings in the magnetic subsystem, increasing the total energy consumption of the hybrid circuit. Moreover, the low electrical output signals of most phenomena severely reduce the noise-limited readout bandwidth and therefore the computing throughput. The magnetoelectric approach, in which electric fields couple to magnetization via strain or other mechanisms, promises a potential solution with much higher energy efficiency and output signal.[75,82,99]. Experimental attempts to demonstrate such converters are actively pursued .

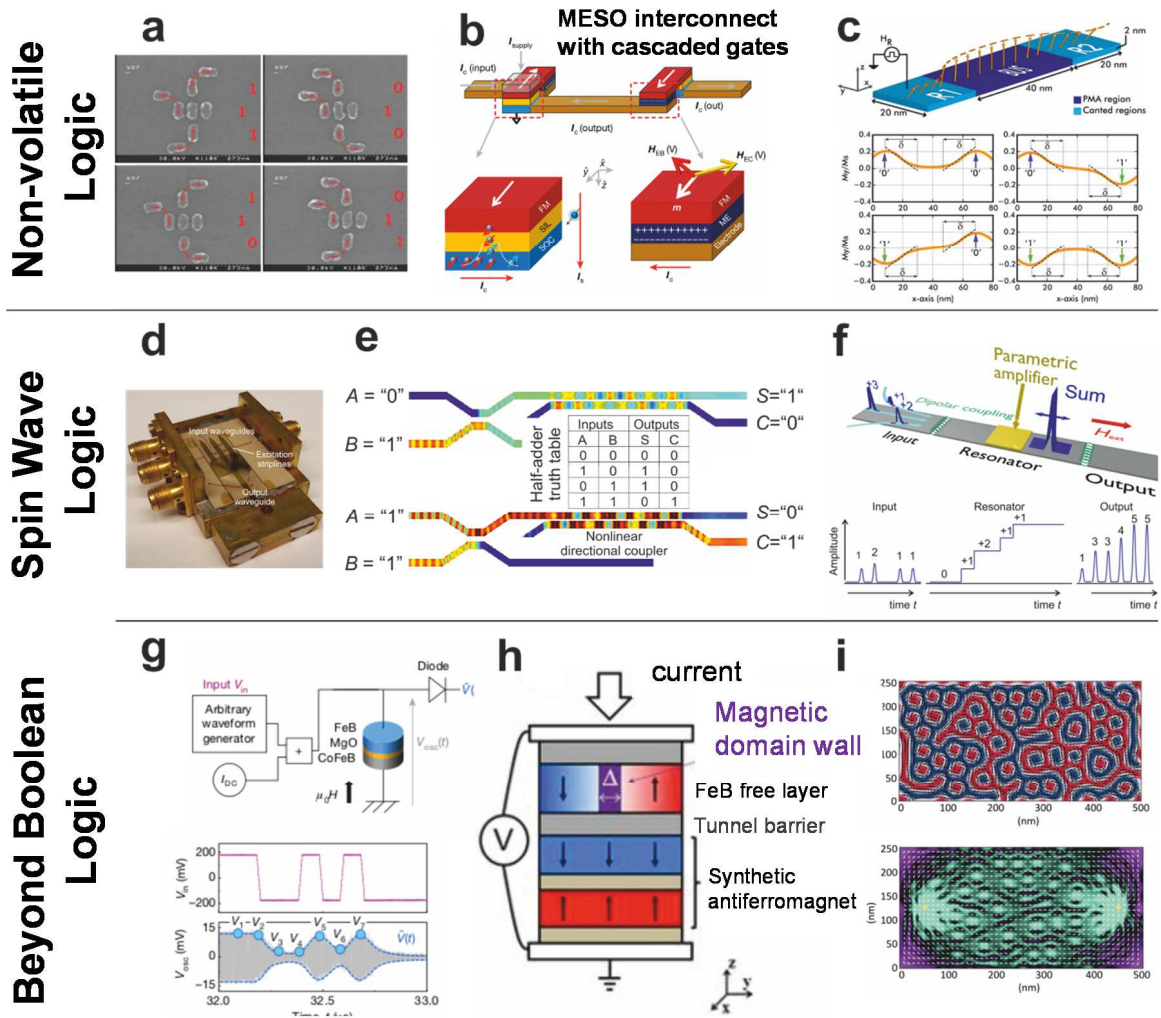

**Fig. 4| Different types of spintronic approaches for logic realizations.** Binary logic based on switching of nanometric magnetic structures (upper panel): (a) Majority gate based on the concept of a magnetic quantum-dot cellular automata using dipolar coupling between nanomagnets[79], (b) Magneto Electric Spin-Orbit (MESO) logic[82] which combines spin-to-charge conversion with magnetoelectric switching for non-volatile, low power logic, (c) Concept of exchange-coupling between bistable magnetic regions which can be used to create majority gates activated by magnetic field pulses.[81] Spin-wave based logic (central panel): (d) Macroscopic prototype of a three-input majority gate based on linear spin-wave superposition[90] (e) All-magnonic half-adder based on nonlinear spin-wave directional couplers[93], (f) Analog spin-wave adder using a parametrically pumped spin-wave resonator for signal storage.[96] Beyond Boolean logic approaches (lower panel): (g) Neuromorphic approach for pattern recognition using the nonlinear properties of spin-torque nano-oscillators[60], (h) Multilevel spin-torque memristor based on a magnetic tunnel junction with a domain wall which can be moved by spin-transfer torque to adjust the resistance[97], (i) Reservoir computing scheme using a magnetic skyrmion network as nonlinear magnetoresistive element.[98]

## Advanced materials, fabrication, and tests

Progress in the areas described above is heavily dependent on functional materials exhibiting high spin polarization, spin filtering, large spin-orbit coupling, high (perpendicular) magnetic anisotropy, and low magnetic damping[100,101]. Moreover, the implementation of novel approaches to manipulate the magnetic state of a device without magnetic fields[100-104] relies on the growth and characterization of materials with atomic scale precision and on the fabrication of complex multilayered structures combining metallic, semiconductor, insulating, and magnetic (both ferro and antiferromagnetic) or non-magnetic materials. This places material science at the heart of any spintronics technology.

*Current status*. Spintronic sensors and MRAM are crucially dependent on interface and spin-filtering effects (Fig. 5a). Standard materials for MTJs are ferromagnetic transition metals (FM), their alloys and the insulator MgO. However, it is becoming a challenge to achieve large magnetoresistance values and reliable switching as the MgO-barrier thickness scales down. The largest TMR at room temperature (604%) was actually reported in a CoFeB/MgO/CoFeB junction a decade ago[107]. In addition, for dense MRAM applications, TMR is not large enough in perpendicular MTJs, requiring amplification at the reading stage, which results in penalties in terms of read speed, ancillary components, and space. Consequently, industry seeks a replacement for MgO and even considers its elimination to return to GMR for some applications. The atomically sharp interfaces that can be achieved with van der Waals heterostructures based on 2D materials (2DM) can help overcome this challenge, and in fact, few-layer graphene has been predicted to act as a spin filter with low resistance.[100] Hexagonal boron nitride (hBN) has also shown potential in Co/hBN/Fe structures with TMR larger than 50%[106]. Another alternative to increase the TMR involves half-metallic ferromagnets, *e.g.*, Heusler alloys, perovskites, rutiles, ferrites and dilute magnetic semiconductors. Heusler alloys have demonstrated 100% spin polarisation at room temperature. Their crystallisation temperature has been recently reduced from >400ºC down to <135ºC[107], allowing their implementation in spintronic devices without interfacial mixing.

In order to enhance the writing efficiency and decrease the energy consumption in STT-MRAM, one could rely on PMA, which leads to a reduced switching current intensity. Graphene can enhance the PMA in FMs such as Co[108]. Another direction is to utilize the spin Hall effect or the inverse spin-galvanic effect in materials with large SOC. Recent reports on FM/topological-insulators (TI) bilayers underline their potential to improve the switching efficiency with SOT. Besides, engineering the SOT through crystalline symmetries can be useful in devices with PMA, as claimed in transition-metal dichalcogenides (TMDC)/FM bilayers[109]. Moreover, strong Dzyaloshinskii-Moriya interaction (DMI) in FM/Graphene, FM/TMDC and FM/TIs can help stabilise topologically protected skyrmions[110].

Spintronic devices based on DMI and 2DMs are associated to new paradigms for data storage and computing. Advances on heterostructures with magnetic 2DMs have resulted in remarkably large magnetoresistances, albeit at low temperature[111, 112] yielding record TMR amplitude in spin-filter van der Waals heterostructures of 19,000%[112] (See Fig.5c) In terms of spin-based computing, graphene can transfer spin information over tens of micrometers for spin-interconnects or reprogrammable magneto-logics.[100] In addition, proximity effects in 2DMs offer a wealth of opportunities not available in bulk materials. Imprinting the spin-valley coupling of TMDCs in graphene leads to spin relaxation anisotropy for room-temperature spin filters and switches[113-115]; magnetic proximity effects in graphene from magnetic EuS and yttrium iron garnet (YIG) have also been observed[116].

Besides, integrating garnets, perovskites and other functional crystalline spintronic materials on silicon CMOS or III-V substrates using direct/buffered epitaxy or wafer bonding could help to achieve high figure-of-merit nonreciprocal CMOS photonics, in-memory computation and high-frequency logic operation capabilities.

*Future challenges.* Sensor and memory applications require a steady improvement in the scaling, the sensitivities and signal to noise ratio. A specific target is to fabricate 10 Gbit MRAM (Fig. 5a) but several challenges must be addressed. Current MTJ stacks are complex and relatively thick (~20nm), hindering the ion-beam etching process and setting a lower limit for the memory-cell pitch. For energy efficient SOT writing, spin-charge interconversion must be improved, while for STT writing, switching currents must be decreased and TMR increased. A crucial challenge relates to engineering interface properties, *e.g.* between new FMs and insulators, TIs and FMs or between 2DMs. Proximity effects or DMI must be thoroughly characterised. Specific tools enabling the growth of complex spintronics heterostructures were assembled such as the Davm tube in Nancy (Fig.5b).

In order to move to high TRLs, high quality nanostructured materials must be obtained on large surfaces. Common spintronic materials such as Co and Cu are increasingly used in conventional electronics, simplifying their integration. This is not the case for Heusler alloys, TIs or 2DMs. Embedding them into spintronic technologies requires the development of large-scale production tools, in order to transfer or directly grow these materials onto 300mm wafers. To incorporate 2DMs into MRAM also requires deposition of magnetic multilayers by PVD, as favored by industry, without degrading the 2DM properties.

Regarding testing and characterization, tools and facilities must be developed to collect specific information associated to spintronic devices for large volume production and quality control. They particularly include techniques to better assess tunnel barrier quality and device reliability.

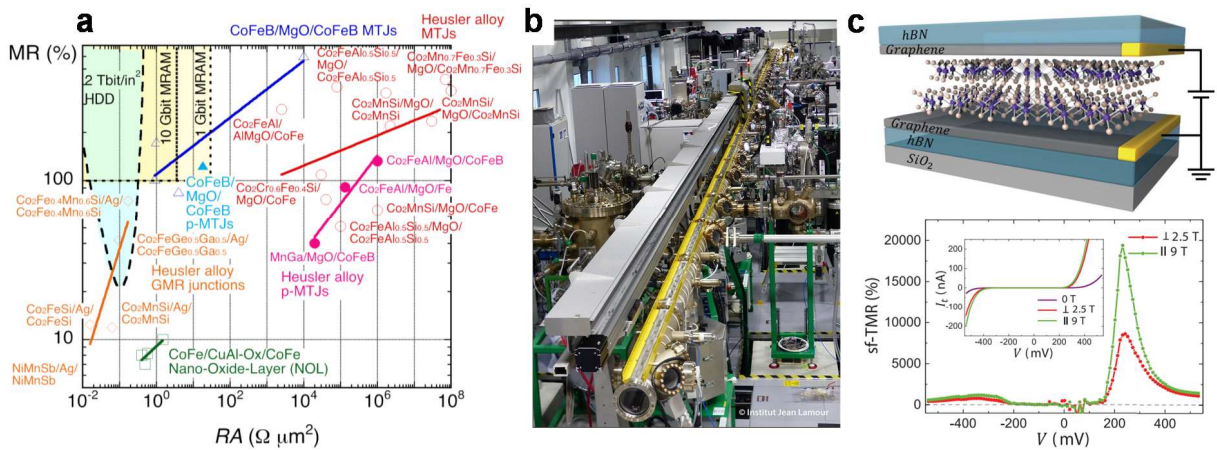

**Fig. 5| Materials and nanofabrication challenges**. (a) Relationship between magnetoresistance and resistance-area product (*RA*) of a series of junctions. The target requirements for 2 Tbit/in$^2$ hard disk drive (HDD) read heads as well as 1 and 10 Gbit MRAM applications are shown. (b) Tube Davm: 30 equipments for growth and characterization are connected under ultra-high vacuum along a 70 meter long Tube. (c) CrI3 spin filter van der Waals heterostructures exhibiting a record magnetoresistance of 19,000% at low temperature (from Ref. 112).

**Modelling and design**

Because of the interplay between structural, magnetic, and electron transport properties at the nanoscale, modelling is of the utmost importance at every stage of the development of spintronic devices. Fundamental models aim at describing the relevant physics at different spatial and time scales and predicting device performance. At the technological level, models support the optimization of devices against industrial specifications and provide tools for circuit designers.

*Current status.* The design of spintronic devices is an intrinsically multiscale problem (see Fig. 6). There is no single model capable of describing all underlying physical processes at all spatial and temporal scales, nor is such a model expected in the foreseeable future. Instead, progress proceeds via an ensemble of theories operating at different levels, each illuminating part of the picture. Some models are well established, whereas others are still evolving: however none is of itself complete: rather they are largely complementary and communicate important information.

Starting from the atomic scale, *quantum-mechanical ab-initio models* using density functional theory[76] have been very successful in predicting, for example, intrinsic magnetic properties central to spintronic functionalities. Together with the non-equilibrium Green's function formalism they are able to predict the transport properties of complex structures. However, these calculations remain idealized and lack many important features significantly influencing spintronic applications, including defects, finite temperature and realistic magnetization dynamics.

Finite size effects and disorder require discrete *classical* spin models[117]. These models correctly describe thermal spin waves and can account for local disorder. They are used to study dynamical magnetic properties in different systems. Coupled to spin transport models, they are suitable for calculating spin currents and spin accumulation at interfaces on the length scale of nanometres, and the dynamic response of the magnetic material.

Larger length scales are the province of *micromagnetics*[118], a mesoscopic theory that has proven very useful due to its ability to reproduce experimental data quantitatively. It is based on the Landau-Lifshitz-Gilbert equation for the magnetization dynamics, and can include phenomena such as spin-transfer, and spin-orbit or Rashba torques. These additional terms introduce unknown parameters, which are sometimes measured experimentally or calculated by ab-initio models. A more proper description for temperature effects is based on the Landau-Lifshitz-Bloch equation[119].

Finally, designing hybrid CMOS/magnetic logic circuits requires integrating the magnetic devices into the standard design suites of microelectronics. Consequently, *compact models* of magnetic devices have been developed[120,121]. Here the physical behavior of the device is converted into an electrical equivalent circuit whose operational characteristics are integrated with the appropriate electrical simulator. Compact models of MTJs are widely used. Some compute the magnetization dynamics,[122] giving high accuracy and particular suitability for analogue functionalities, such as sensing or RF communication. Others consider bi-stable magnetization states and calculate switching times triggered by current pulses.[121] This much faster approach allows the simulation of larger circuits, such as memory arrays.

Despite continuous advances in modelling, key aspects need to be accelerated to match the fast progress of experiments and to meet the industrial demands. The following priorities are identified:

*Bridging gaps.* Since no single model can describe all physical effects in a complete picture, the coupling of different approaches is paramount and a more elaborate hierarchical multi-scale description is required[122]. Here atomistic on-site parameters (anisotropy, exchange, Dzyaloshinkii-Moriya interactions) are evaluated with ab-initio theories. Atomistic models then evaluate the temperature dependence of macroscopic parameters, which serve as input for large-scale micromagnetic simulations. This atomistic/micromagnetic interface should be thoroughly investigated, especially including proper characterization of cell-size effects. Finally, micromagnetic simulations are used to test the accuracy of compact models.

*Improving current models.* Current model approaches are constantly evolving and undergoing improvement in terms of accuracy, flexibility and efficiency. Ab initio models need to move towards larger-scale simulations with thousands of atoms and novel approaches beyond the local density approximation need to be developed, also considering disordered magnetic states at finite temperatures. With respect to discrete atomistic spin models, the main challenge is to introduce quantum effects and correlations in the classical description. The development of mesoscopic approaches, such as micromagnetics, must include coupling to other models of physical phenomena that affect magnetization. For example, a model coupling both charge and spin transport self-consistently with magnetization dynamics is required[123,124]. Finally, the integration of compact and efficient electrical circuit models of spintronic devices into the standard design suites of CMOS-integrated electronics using a PDK (Process Design Kit) is essential.

*Developing models to investigate new phenomena.* The continual discovery of new spin-dependent phenomena requires prompt development of theoretical models and tools capable of describing the physics of new materials and systems. We highlight the following:

- *Ultrafast all-optical switching*[125], where dynamic magnetization processes are triggered by ultrafast laser excitation. This involves complex interactions of various subsystems (electrons, phonons, magnons) with the laser pulse.

- *Terahertz* spintronics[31], where interaction of the THz electric pulse with a magnetic medium efficiently excites magnetization dynamics and is far from understood.

- *Spin* caloritronics[24] requires models to describe the interplay between heat, charge and spin currents for possible energy harvesting.

- *Voltage and electric field switching*[126] requires models capable of predicting the complicated dynamical influence of electric fields on ferromagnets.

- *Antiferromagnetic spintronics*[127] requires a large-scale model description for simulation of nanodevices, which is as yet in the early stages of development.

*Integration and interconnectivity.* A common strategic view transverse to the above aspects consists of working towards standardization of modelling approaches. For instance, compact models can be associated with a "model card", which provides the technology parameters and their typical variations. The next step will be to evolve the cards along with developing technology, possibly using automatic characterization tools and fitting algorithms. This process is expected, similarly to that developed by the Compact Model Coalition for CMOS electronics, to boost interconnectivity among different sources, thereby increasing the efficiency of feedback between industrial modelling and device design

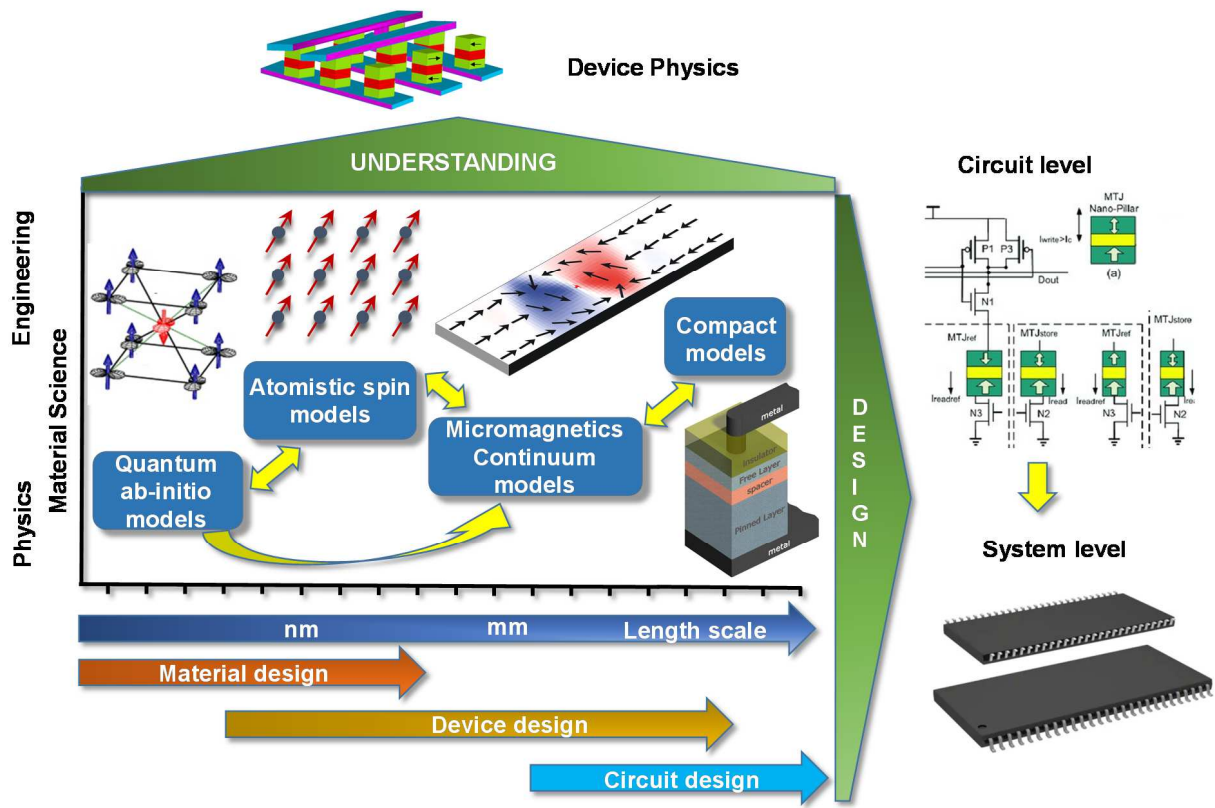

**Fig. 6| Modeling approaches used in spintronics.** The figure illustrates the need for different models to bridge atomistic insight into fundamental physics and materials science with reliable predictions and designs at the device and circuit levels.

**Conclusions and outlook**

Expanding beyond well-established sensor applications, spin-electronics can bring major advantages to the microelectronic industry. In this review, we have highlighted four major areas of impact with associated medium-term technological developments.

MRAMs have a proven potential to replace existing memory technologies for applications where the combination of non-volatility, speed, and endurance is key. STT-MRAM is a high density and scalable technology, which is well suited for high performance computing applications at the last level caches. STT-MRAM is not suitable for faster caches, for which SOT-RAM could be a solution. Electric field control of the magnetization, possibly in combination with spin-charge interconversion, as well as optical switching of the magnetization (photonics/spintronics coupling) could further extend the range of spintronic memories beyond-CMOS technologies.

Magnetoresistive based sensing is increasing its market share in applications where higher output and SNR, good thermal stability, compatibility with CMOS integration, reduced cost and small footprints (< 1mm$^2$) are required. With the transition to electrical vehicles (cars and grids), increasing need to use renewable energy sources (solar and wind generators and associated invertors), industry 4.0 and IoT concepts, the need for smarter and more advanced magnetic sensors is growing.

Meanwhile, RF spintronics is reaching maturity, both in terms of material developments, nanofabrication (single devices and arrays), and control of the dynamical modes as a function of the device configuration, up to the first system-level RF circuits. RF spintronics bears a huge potential to provide novel, low cost, low power, more-than-Moore and/or beyond-CMOS approaches for applications in IoT, 5G, electronic smart systems or flexible electronics. First CMOS integrated demonstrators are expected in the next 3-5 years.

Spintronic logic is promising for ultralow power electronics due to the low intrinsic energy needed to manipulate nanomagnets and magnetic excitations such as magnons at the nanoscale. A large number of different logic concepts have been proposed although at the current stage of development only very few of them allow for the design of complex circuits. Hybrid systems combining spintronics and CMOS are required to validate these concepts at system level and benchmark them against traditional CMOS approaches. Ultimately, the magnetoelectric effect promises to achieve necessary energy efficiency and signal levels if the amplitude of the effect can be increased and its scalability to the nm-GHz range demonstrated. Spintronic devices are also attractive for the realization of novel and promising non-Boolean computing approaches that do not necessarily require ultralarge-scale integration, such as neuromorphic computing inspired by the functionality of the human brain.

As has been the case for the industrialization of devices based on GMR, TMR, and STT, progress along these routes will be achieved by developing novel materials, functional heterostructures, and device architectures. Simulations tools that bridge the gap between the atomistic, device, and system levels may significantly speed up the translation of results from basic research into microelectronic technologies.

The launching of volume production of MRAM marks the acceptation of the spintronics technology by the microelectronics industry. This major step is now past and spintronics is becoming a mainstream technology. This evolution of the state of mind will allow for faster integration off the new breakthrough discoveries in the field made in the past ten years, many of which are bearing perspectives of applications in fields as diverse as ultra-low power electronics, IoT, RF communication, energy harvesting, artificial intelligence, cryoelectronics, quantum engineering… Electron spin will likely play a major role in the future of microelectronics.

**Additional information**

The authors declare no competing financial interests.